# Magnetic Proximity Effect in Pt/CoFe$_2$O$_4$ Bilayers

Walid Amamou,[1*] Igor V. Pinchuk,[2*] Amanda Hanks,[3,4] Robert Williams,[3] Nikolas Antolin,[4] Adam Goad,[2,5] Dante J. O'Hara,[1] Adam S. Ahmed,[2] Wolfgang Windl,[4] David W. McComb,[3,4] and Roland K. Kawakami[1,2**]

[1]*Materials Science and Engineering, University of California, Riverside, CA 92521*
[2]*Department of Physics, The Ohio State University, Columbus, OH 43210*
[3]*Center for Electron Microscopy and Analysis, The Ohio State University, Columbus, OH 43210*
[4]*Department of Materials Science and Engineering, The Ohio State University, Columbus, OH 43210*
[5]*Department of Physics, University of Maryland, Baltimore County, MD 21250*

**Abstract**

We observe the magnetic proximity effect (MPE) in Pt/CoFe$_2$O$_4$ bilayers grown by molecular beam epitaxy (MBE). This is revealed through angle-dependent magnetoresistance measurements at 5 K, which isolate the contributions of induced ferromagnetism (i.e. anisotropic magnetoresistance) and spin Hall effect (i.e. spin Hall magnetoresistance) in the Pt layer. The observation of induced ferromagnetism in Pt via AMR is further supported by density functional theory calculations and various control measurements including insertion of a Cu spacer layer to suppress the induced ferromagnetism. In addition, anomalous Hall effect measurements show an out-of-plane magnetic hysteresis loop of the induced ferromagnetic phase with larger coercivity and larger remanence than the bulk CoFe$_2$O$_4$. By demonstrating MPE in Pt/CoFe$_2$O$_4$, these results establish the spinel ferrite family as a promising material for MPE and spin manipulation via proximity exchange fields.

* equal contributions
**email: kawakami.15@osu.edu

Spin manipulation inside a nonmagnetic (NM) material using internal effective fields (spin orbit or exchange) is a very promising avenue toward the realization of next generation spintronic devices (spin transistors, magnetic gates, etc.) [1,2]. In particular, magnetic proximity effect (MPE) at the interface of a NM spin channel and a ferromagnetic insulator (FMI) is of great importance for generating exchange fields and induced ferromagnetism in the NM layer. Recently, spin manipulation by MPE has been realized in experiments that modulate spin currents in graphene on YIG (yttrium iron garnet) [3,4]. In addition, proximity exchange fields induced by FMI have been observed for graphene and monolayer transition transition metal dichalcogenides [5,6].

Among FMIs, members of the spinel ferrite family ($MFe_2O_4$, with M=Co, Ni, etc.) are attractive because their magnetic properties can be tuned by alloy composition [7,8] as well as epitaxial strain [9-11]. In particular, $CoFe_2O_4$ (CFO) is a hard ferrimagnetic insulator which exhibits high Curie temperature (728 K), spin-filtering properties [12-14], magneto-electric switching [15], and is readily integrated with practical spintronic materials (MgO, Fe, Cr). Unfortunately, all previous experiments have failed to observe MPE using CFO. To test for MPE in NM/FMI systems, Pt is widely used as the NM material due its closeness to fulfilling the Stoner criteria and thus allowing it to become ferromagnetically ordered at the interface with the FMI. Initial studies of Pt/CFO grown by pulsed laser deposition (PLD) utilized magnetotransport measurements and observed no MPE in the Pt layer [16]. Subsequent transport and element-specific magnetization measurements by x-ray magnetic circular dichroism (XMCD) also found no evidence for induced ferromagnetism in the Pt layer [17-20] contributing to a growing consensus that CFO cannot be used to obtain MPE. However, because the length scale of exchange interactions is on the order of angstroms, it should depend on the atomic structure and alternative growth methods may be needed for its realization.

In this Letter, we utilize molecular beam epitaxy (MBE) to synthesize Pt/CFO bilayers and observe induced ferromagnetism (i.e. MPE) in the Pt layer at 5 K. This is revealed through angle-dependent magnetoresistance measurements, which isolate the contributions of induced ferromagnetism (i.e. anisotropic magnetoresistance, AMR) and spin Hall effect (i.e. spin Hall magnetoresistance, SMR) in the Pt layer. The observation of induced ferromagnetism in Pt via AMR is further supported by density functional theory (DFT) calculations and various control measurements including insertion of a Cu spacer layer to suppress the induced ferromagnetism. In addition, anomalous Hall effect (AHE) measurements show an out-of-plane magnetic hysteresis loop of the induced ferromagnetic phase with larger coercivity and larger remanence than the bulk CFO. By demonstrating MPE in Pt/CFO, these results establish the spinel ferrite family as a promising material for MPE and spin manipulation via proximity exchange fields.

The CFO thin films are grown on MgO(001) substrates by reactive MBE (details in Supplementary Materials) and are characterized by reflection high-energy electron diffraction (RHEED), atomic force microscopy (AFM), x-ray diffraction (XRD), and high angle annular dark field scanning transmission electron microscopy (HAADF STEM). Figure 1a shows a RHEED pattern of CFO(40 nm)/MgO(001) taken along the [110] in-plane direction. The image displays streaky and sharp diffraction maxima, indicating a flat and single crystal surface. This is confirmed by AFM, which exhibits very smooth morphology over large areas (Figure 1b) with an rms roughness of 0.06 nm over a 10 μm x 10 μm area of a 40 nm CFO film. The crystallinity is confirmed by θ-2θ XRD scans on Pt(1.7 nm)/CFO(40 nm)/MgO(001), which exhibit clear MgO(002) and CFO(004) peaks and no other phases within the scan range (Figure 1c). A clearly separable CFO (004) (inset, Figure 1c) peak gives a perpendicular lattice constant of 8.365 Å, indicating a CFO film under slight tensile strain compared to bulk CFO lattice constant of 8.392 Å[10]. Finally, a cross-sectional HAADF STEM image shows an atomically sharp interface between the MgO substrate and CFO thin film with an epitaxial relationship of $[100]_{MgO}$ // $[100]_{CFO}$ and $[010]_{MgO}$ // $[010]_{CFO}$, as indicated in Figure 1d. The appearance of the lattice varies across the CFO thin film, switching between a cubic appearance and that which resembles a spinel structure. We believe this contrast variation is due to changes in the degree of inversion, $\lambda$. In $AB_2O_4$ spinels, $\lambda$ adopts values between 0 (normal) and 1 (inverse), and is equal to the fractional occupancy of the trivalent $B^{3+}$ cation on the tetrahedral A-site sub-lattice.

Bulk magnetic properties of a Pt(1.7 nm)/CFO(40 nm) sample are measured at 5 K by vibrating sample magnetometry (VSM) (Figure 1e). An in-plane hysteresis loop taken along the [100] axis (red curve) has a coercivity $\mu_0 H_C$ of 0.34 T, saturation field of $\mu_0 H_s$ of 3.3 T and a remanence ratio $M_R/M_S$ of 0.1. An out-of-plane hysteresis loop (black curve) has similar characteristics, coercivity $\mu_0 H_C$ of 0.18 T, saturation field of $\mu_0 H_S$ of ~4.25 T and a remanence ratio $M_R/M_S$ of 0.06.

To detect MPE in Pt/CFO, we perform magnetotransport measurements that are sensitive to the presence of magnetization within the Pt layer. In ferromagnets, two well-known phenomena are the anomalous Hall effect (AHE) which is sensitive to the out-of-plane magnetization, and the anisotropic magnetoresistance (AMR) which is sensitive to the orientation of magnetization relative to the current direction. With induced magnetization in the Pt layer along unit vector $\mathbf{m_{Pt}}$, these appear in the longitudinal and transverse resistivities as:

$$\rho_{xx} = \rho_0 + \Delta\rho_{AMR} m^2_{Pt,j}$$
$$\rho_{xy} = \Delta\rho_{AMR} m_{Pt,t} m_{Pt,j} + \rho_{AHE} m_{Pt,n} \qquad (1)$$

where $m_{Pt,n}$, $m_{Pt,jt}$ $m_{Pt,t}$ are the out-of-plane (**n**), in-plane along current (**j**), and in-plane transverse to current (**t**) components of the Pt magnetization unit vector (see Figure 3(a)), $\rho_0$ is the background resistivity of Pt, and $\Delta\rho_{AMR}$ and $\rho_{AHE}$ are the MPE-induced AMR and AHE, respectively. In addition to AHE and AMR, a recently discovered pure spin current effect based on the spin Hall effect in Pt and interfacial spin scattering at the FMI interface generates additional contributions to $\rho_{xx}$ and $\rho_{xy}$ given by:

$$\rho_{xx} = \rho_0 + \Delta\rho_1 \mathbf{m}^2_{CFO,t}$$
$$\rho_{xy} = -\Delta\rho_1 \mathbf{m}_{CFO,t}\mathbf{m}_{CFO,j} + \Delta\rho_2 \mathbf{m}_{CFO,n} \qquad (2)$$

where $m_{CFO,j}$, $m_{CFO,t}$, $m_{CFO,n}$ are components of the magnetization unit vector in the FMI, $\Delta\rho_1$ is known as the spin Hall magnetoresistance (SMR), and $\Delta\rho_2$ is the spin Hall anomalous Hall-like signal (SH-AHE). The SMR stems from the reflection of spin current (generated by spin Hall effect) from the FMI interface which is subsequently converted to a charge current through the inverse spin Hall effect (ISHE) [21]. The SH-AHE stems from reflection of the spin current at the FMI interface, where an out-of-plane component of FMI magnetization rotate the spin orientation of the spin current and generate a transverse voltage via ISHE. Finally, in addition to the AMR and SMR effects, one must also consider the ordinary magnetoresistance (OMR) and ordinary Hall effect (OHE) that occur due to the presence of Lorentz forces acting on charge carriers in a magnetic field. With the possibility of many different effects (OMR, AMR, SMR, OHE, AHE, SH-AHE) contributing to $\rho_{xx}$ and $\rho_{xy}$, a systematic approach is essential for identifying the presence of MPE.

We begin with angle-dependent (AD) magnetoresistance measurements to separate the contributions from AMR, SMR, and OMR (see Supplementary Materials for measurement details). Considering that AMR depends on the *j*-component of magnetization (equation 1) and SMR depends on the *t*-component of magnetization (equation 2), the two effects can separated by rotating the magnetization within different planes. For ADAMR, the relevant angular scan is in the *n-j* plane, where γ is defined as the angle measured from the normal axis (*n*) (see Figure 2a) while ADSMR does not depend on γ. For ADSMR, the relevant angular scan is in the *n-t* plane, where β is defined as the angle measured from the normal axis (*n*) (see Figure 2b) while ADAMR does not depend on β. Finally, the contribution from OMR has the same functional form as AMR (i.e. depends on γ), but the OMR can be determined independently. OMR in most materials has a larger resistance when the magnetic field is perpendicular to the current (γ = 0°) as compared to parallel to the current (γ = 90°), and we have verified this for our Pt films on MgO(001) substrates as well (inset Figure 2c). To determine if the Pt/CFO system exhibits MPE, we therefore perform a γ-scan to look for the presence of AMR. Figure 2c shows clearly the presence of angle-dependent MR with lower resistivity for γ = 0° and higher resistivity for γ = 90°. Because this cannot be explained by OMR (opposite polarity) and the γ-scan is insensitive to SMR, it is clear evidence for AMR and induced ferromagnetism in the Pt layer. By comparison, Figure 2c inset shows γ scans of Pt/MgO and Cu/CFO with both displaying OMR oscillations. This is the strongest evidence for MPE in Pt/CFO in our study. Such an AMR signature has never been observed in previous studies of Pt/CFO, but it has been previously reported for other Pt/FMI systems and is accepted as the most reliable test among transport measurements for MPE [22-24]. We also perform the β-scan and observe SMR with similar magnitude as reported in previous studies [16-18,20] of Pt/$CoFe_2O_4$ (Figure 2d).

We also perform Hall measurements to further support the presence of induced ferromagnetism in the Pt layer. Measurement of $R_{xy}$ at 5 K for the Pt (1.7 nm)/CFO(40 nm) sample (Figure 2e red curve) exhibits a nonlinear hysteretic signal commonly associated with AHE (and thus ferromagnetism) and a linear OHE background. The absence of such nonlinear features in Cu (8 nm)/CFO(40 nm) and Pt (5 nm)/MgO control samples rule out magnetic fringe fields or magnetic impurities in Pt, respectively, as the origin of the hysteresis loop in the Pt/CFO sample. It is interesting to note that after the linear OHE contribution is subtracted (Figure 3a, red curve), the remaining hysteretic signal of Pt/CFO shows higher coercivity and substantially larger remanence ratio than the out-of-plane hysteresis loop of bulk CFO (Figure 1e). These different interfacial magnetic properties might be due to the exchange interaction between the CFO and Pt moments or a surface magnetic anisotropy at the CFO/Pt interface.

The Hall resistivity $\rho_{xy}$ in Figure 3a can come from two sources as shown in Equations (1) and (2): MPE-AHE and SH-AHE. To further distinguish between these mechanisms, we insert a 2 nm Cu spacer between Pt and CFO layers. We utilize Cu because it has (1) filled d-shells which prevents induced ferromagnetism, and (2) a long spin diffusion length (hundreds of nm), which makes it transparent to spin currents. Therefore, a Cu spacer should suppress induced magnetism in Pt while leaving SH-AHE and SMR unaffected (except for shunting effects). The Hall measurement of the Pt/Cu/CFO sample (Figure 4a, blue curve) indeed shows a reduction in magnitude compared to Pt/CFO, which can come from a loss of induced ferromagnetism and/or a reduction of the overall signal by shunting.

Further insight is gained through ADMR measurements comparing Pt/Cu/CFO and Pt/CFO. Figure 3b shows a plot of the resistivity ratio $\Delta\rho_1/\rho_0$ as a function of angle *β* for Pt/CFO and Pt/Cu/CFO at 5 K. Notably, the ADSMR signal yields nearly the same modulation as the direct contact, with the $\Delta\rho_1/\rho_0$ magnitude practically unchanged. This suggests that the shunting through the 2 nm Cu layer is minimal. Furthermore, Figure 3c compares the ADAMR signal in the Pt/Cu/CFO bilayer (blue curve) and the Pt/CFO bilayer (red curve). Indeed, the Pt/Cu/CFO sample does not show any modulation arising from the AMR. Although this does not prove an absence of AMR (because it might be getting cancelled by OMR), it nevertheless shows a significant reduction compared to the Pt/CFO sample. This reduction in AMR provides additional evidence for induced ferromagnetism in Pt/CFO and suggests that the reduction in AHE in Pt/Cu/CFO is primarily due to the reduction of induced ferromagnetism.

Finally, we investigate the MPE in Pt/CFO using density functional theory calculations. Using the Vienna Ab-initio Simulation Package (VASP), we relax cubic CFO cells consisting of 5 F.U. with the Co atoms placed only on the tetrahedral sites, ordered octahedral sites, random octahedral sites, and a combination of random tetrahedral and octahedral sites [25]. These calculations indicate CFO favors Co occupancy of the octahedral sites to minimize structural energy with random distribution and no preference for ordering. Using the lowest energy CFO structure, we construct calculation cells consisting of either one or two cubic CFO cells topped with either 4 (~1 nm) or 8 (~2 nm) layers of Pt with an (001)/(001) interface and at least 10 Å of vacuum. After comparing several Pt positions over the CFO cells, the positions directly over the cation sites are found to be energetically favorable, as shown in Fig. 5b. Finally, we relax the interface calculation cells in the plane of the interface to determine the electronic and magnetic structure of the CFO/Pt interface. All calculations are performed using generalized gradient approximation pseudopotentials in the formulation of Perdew et al.; on-site corrections for Coulomb interactions (DFT+U) are applied based on previous DFT work [26,27]. Calculations utilize spin polarization and Monkhorst-Pack k-point meshes consisting of 7 points in periodic directions and a single point in vacuum directions.

Figure 4a displays the layer-averaged magnetic moments on Pt atoms in layers adjacent to the CFO/Pt interface. The large error bars in the interface layer comes from site-specific variations between the Pt atoms that we will discuss in the Supplementary Materials (section 3). We observe decreasing moment on Pt atoms as a function of distance from the CFO/Pt interface, indicating that the presence of the Pt moments should be due to induced magnetism from the CFO substrate. The magnetic effect does not persist after the first two Pt planes, giving a length scale for the proximity effect. Figure 4b displays the magnetic moments on Pt atoms in the 8-layer calculation cell with isosurfaces at 0.0025 $\mu_B$, showing that the induced negative moments (red) are only present in the first few atomic planes. To investigate the nature of the induced moments at the interface, we examine the orbital density of states of interfacial Pt atoms on top of various CFO sites (e.g. octahedral Fe, tetrahedral Fe, oxygen, octahedral Co). In particular, as shown in Figure 4c, we observe a strong spin asymmetry (bold line) and induced moment in the density of states of the $dz^2$ orbitals of Pt on top of octahedral Fe, while the other orbitals show much less spin asymmetry. Furthermore, a comparison with the $d$-DOS of Pt on top of other CFO sites and an examination of the $d$-DOS of Fe suggests that magnetism is primarily induced in the $dz^2$ orbital of the Pt atoms by moments in the $dxy$, $dyz$, and $dxz$ orbitals of Fe atoms located no more than one layer below the CFO/Pt interface (see section 3 of the Supplementary Material for details).

In conclusion, we obtain strong evidence for the presence of induced ferromagnetism (i.e. MPE) in Pt/CFO at 5 K including the first measurement of ADAMR in this system. Studies of Hall resistivity, ADSMR, insertion of Cu spacers, and DFT calculations provide additional evidence for MPE. Additional measurements at 300 K indicate the possibility of induced ferromagnetism, but the results are less conclusive due to the absence of a definitive ADAMR signal (see Supplementary Materials). Nevertheless, the observation of MPE in Pt/CFO opens the door to utilizing the family of spinel ferrites for MPE and spin manipulation via proximity exchange fields for novel spintronic devices.

## Acknowledgments

We would like to thank Felix Casanova for insightful discussions. Primary funding for this research was provided by the Center for Emergent Materials: an NSF MRSEC under award number DMR-1420451. NA and WW acknowledge computational support by the Ohio Supercomputer Center under Grant No. PAS0072, and partial funding from AFOSR, award no. FA9550-14-1-0322.

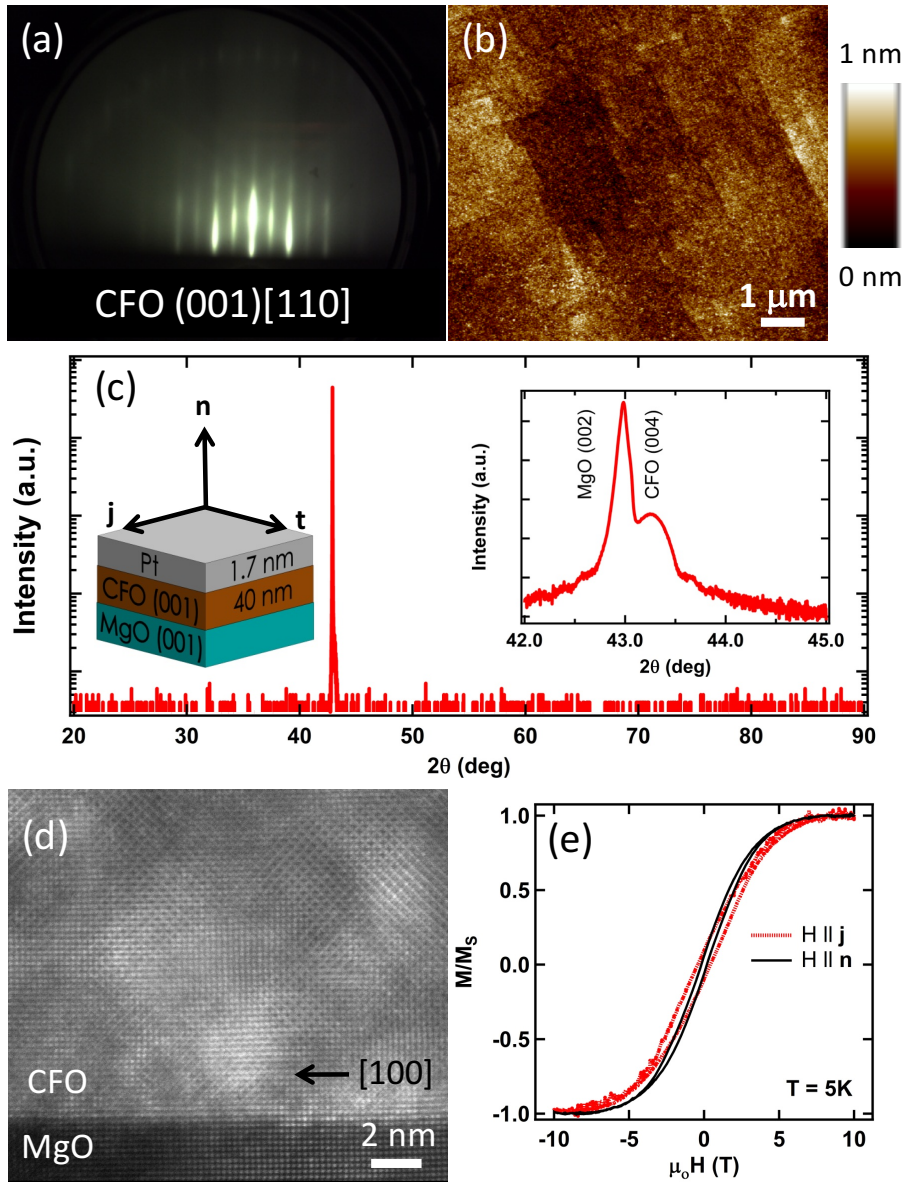

**FIG. 1.** (a) RHEED image of 40 nm CFO film grown on MgO (001) substrate, taken along the [110] in-plane direction. (b) Representative atomic force microscopy image taken over a 10 μm x 10 μm scan size with rms roughness of 0.14 nm. (c) θ-2θ x-ray diffraction scan of Pt(1.7 nm)/CFO(40 nm)/MgO heterostructure. Inset: A fine scan showing a clear CFO (004) peak. (d) HAADF STEM micrograph of a CFO film on an MgO substrate. (e) Magnetic hysteresis loop of of Pt(1.7 nm)/CFO(40 nm)/MgO for both in plane (red) and out-of-plane (black) applied magnetic field.

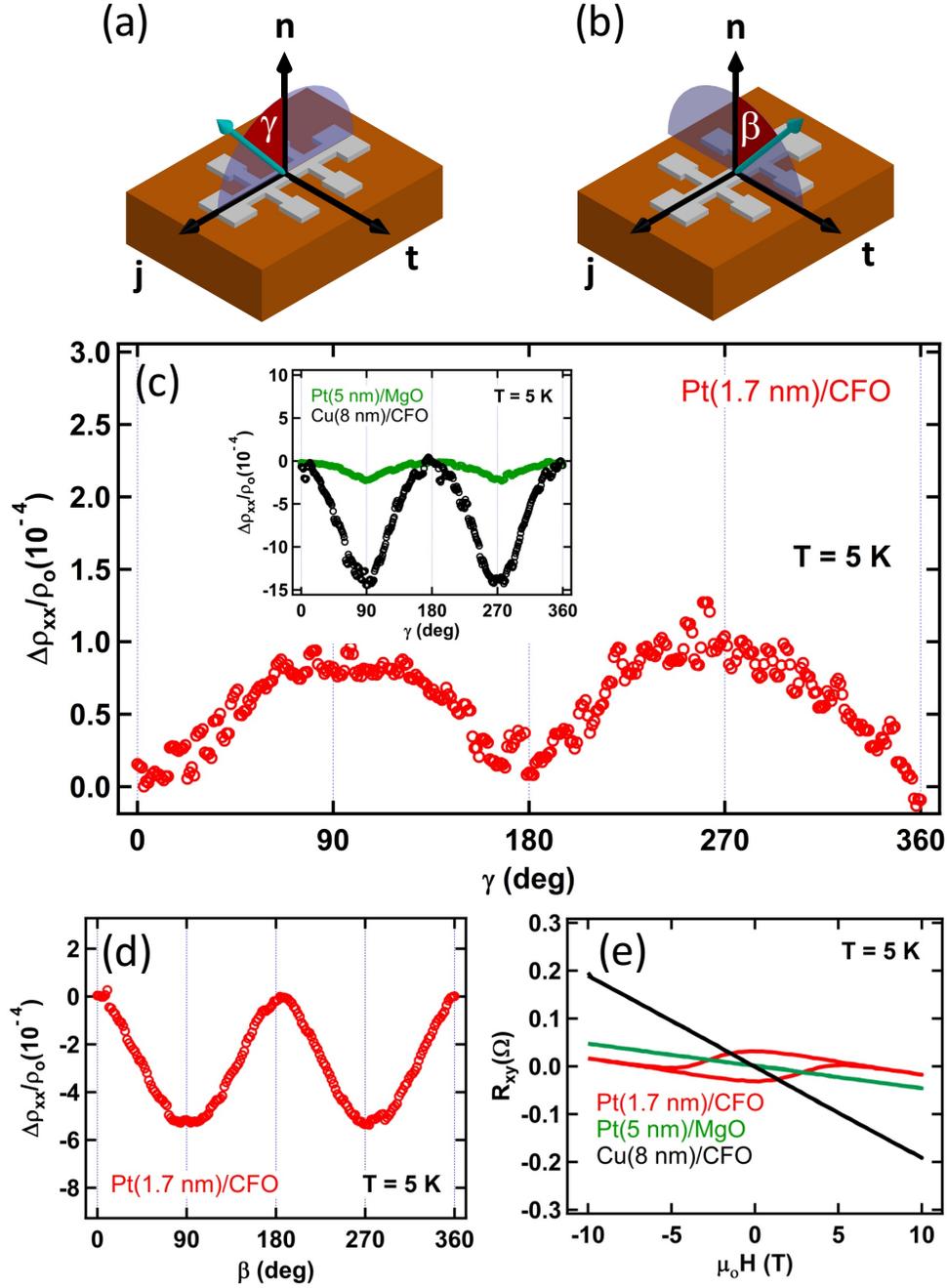

**FIG. 2.** (a) Measurement geometry for ADAMR (γ scan in n-j plane). (b) Measurement geometry for ADSMR (β scan in n-t plane). (c) γ dependence of $\Delta\rho_{xx}/\rho_0$ of Pt/CFO taken with $\mu_0 H$ = 10 T, showing the presence of AMR. (d) β dependence of $\Delta\rho_{xx}/\rho_0$ of Pt/CFO taken with $\mu_0 H$ = 10 T, showing the presence of SMR. (e) Hall resistance for Pt(1.7 nm)/CFO (red curve), Pt(5 nm)/MgO (green curve), and Cu(8 nm)/CFO (black curve). All measurements are taken at T = 5 K.

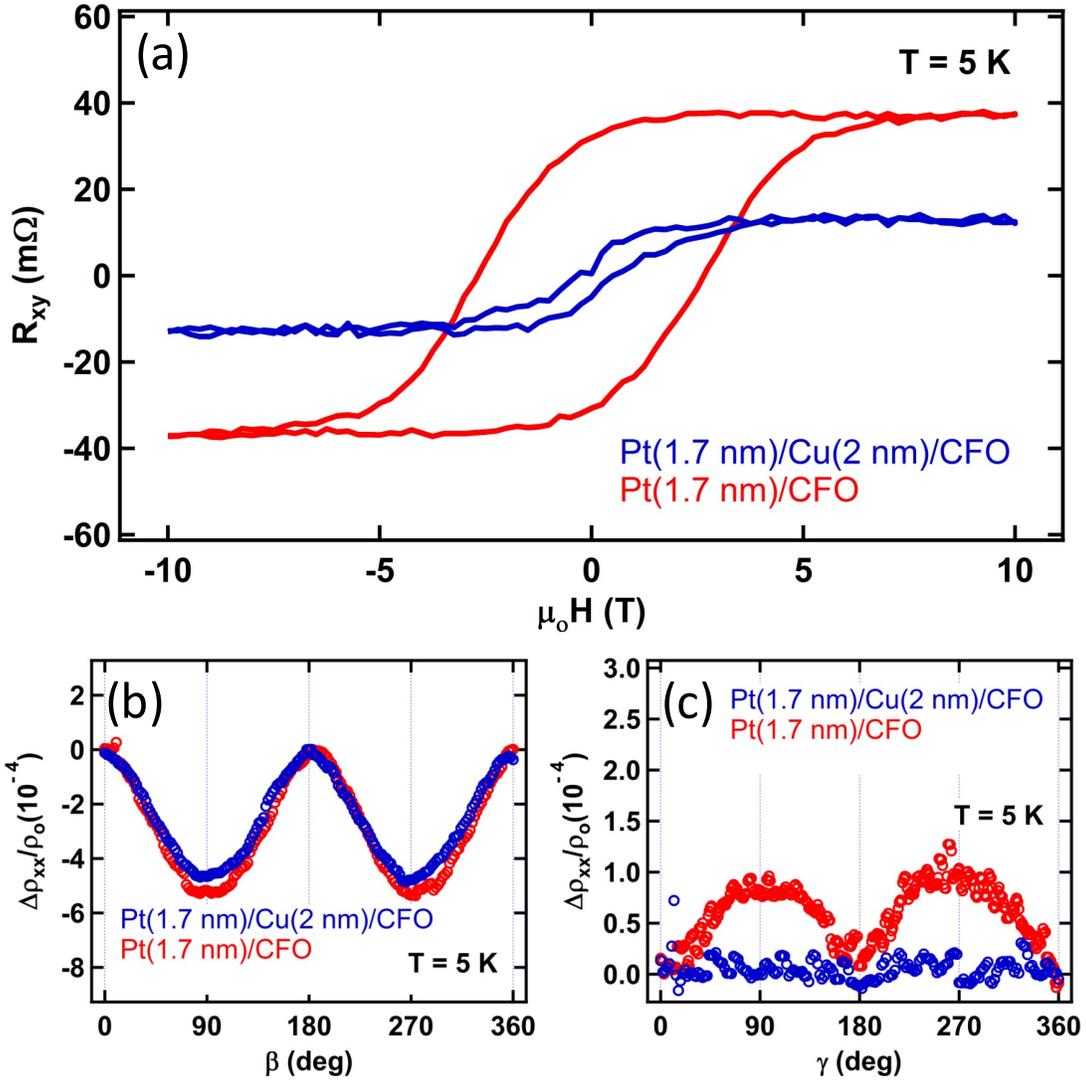

**FIG. 3.** (a) Hall resistivity of Pt(1.7 nm)/CFO (red curve) and Pt(1.7 nm)/Cu(2 nm)/CFO (blue curve) with linear OHE background subtracted. Addition of the Cu spacer heavily suppresses the magnitude and coercivity. (b) ADSMR scans of Pt/CFO (red curve) and Pt/Cu/CFO (blue curve), (c) ADAMR scans of Pt/CFO (red curve) and Pt/Cu/CFO (blue curve). All measurements are taken at T = 5 K.

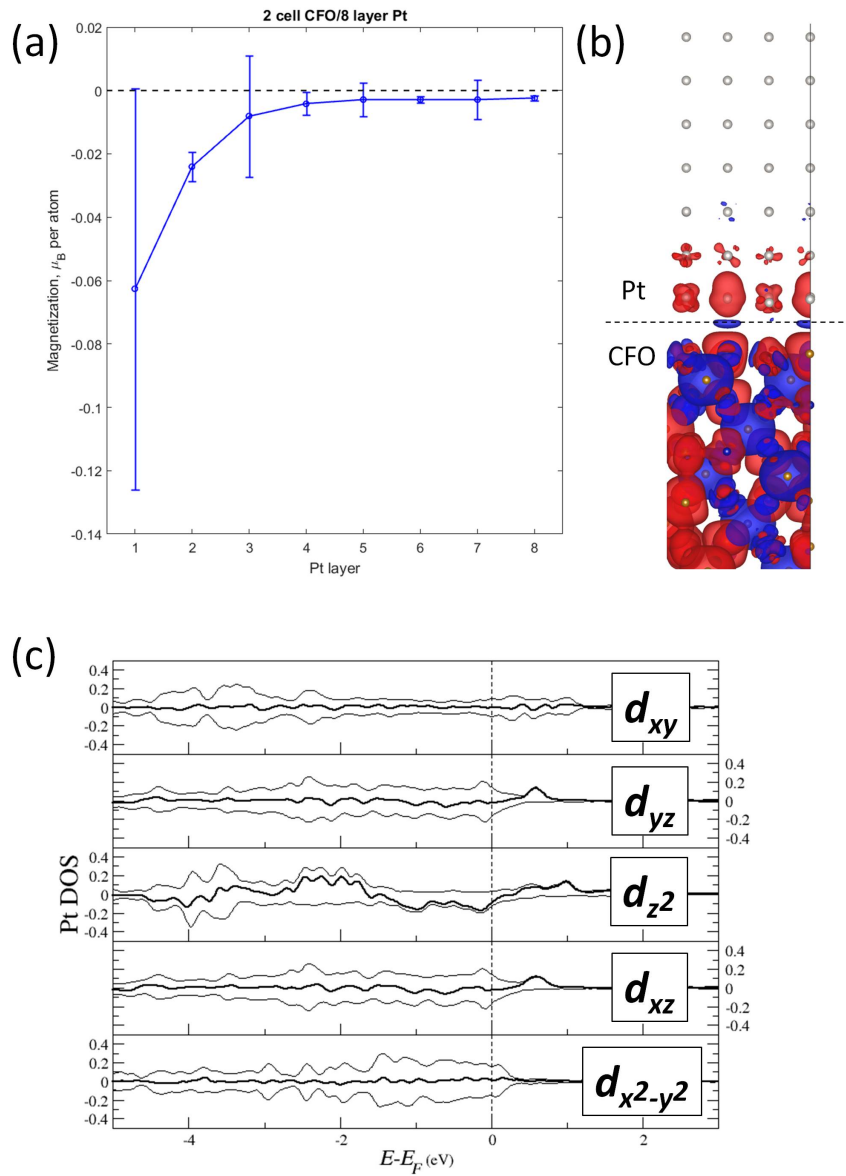

**FIG. 4.** (a) Average magnetic moment per Pt atom in $\mu_B$ for Pt layers adjacent to the CFO/Pt (001)/(001) interface, calculated with DFT. Layer averages decrease sharply to zero in 1-2 layers from interface. (b) Magnetic moment isosurfaces at 0.0025 $\mu_B$ for Pt atoms in the 8-layer calculations cell. Strong negative moments (red) are observed only in the layers closest to the interface, diminishing to zero by the fourth layer of Pt atoms. (c) Orbital-resolved $d$-DOS of a Pt atom located directly over a magnetic Fe atom at CFO/Pt interface. The thin lines are spin up (positive values) and spin-down (negative values) DOS, and the bold line is their sum.

**References**

[1] H. Haugen, D. Huertas-Hernando, and A. Brataas, *Physical Review B* **77**, 115406 (2008).

[2] H. X. Yang, A. Hallal, D. Terrade *et al.*, *Physical Review Letters* **110**, 046603 (2013).

[3] S. Singh, J. Katoch, T. Zhu *et al.*, *Phys. Rev. Lett.* **118**, 187201 (2017).

[4] J. C. Leutenantsmeyer, A. A. Kaverzin, M. Wojtaszek *et al.*, *2D Materials* **4**, 014001 (2017).

[5] P. Wei, S. Lee, F. Lemaitre *et al.*, *Nat. Mater.* **15**, 711 (2016).

[6] C. Zhao, T. Norden, P. Zhang *et al.*, *Nat Nano* **advance online publication** (2017).

[7] T. N. Koichiro Suzuki, Yohtaro Yamazaki, *Japanese Journal of Applied Physics* **27**, 361 (1988).

[8] J. A. Moyer, C. A. F. Vaz, E. Negusse *et al.*, *Physical Review B* **83**, 035121 (2011).

[9] Y. Suzuki, G. Hu, R. B. van Dover *et al.*, *Journal of Magnetism and Magnetic Materials* **191**, 1 (1999).

[10] G. Hu, J. H. Choi, C. B. Eom *et al.*, *Physical Review B* **62**, R779 (2000).

[11] K. U. H. Yanagihara, M. Minagawa, E. Kita, N. Hirota, *Journal of Applied Physics* **109**, 07C122 (2011).

[12] A. V. Ramos, Universite Pierre et Marie Curie (Thesis), 2008.

[13] A. V. Ramos, M.-J. Guittet, and J.-B. Moussy, *Applied Physics Letters* **91**, 122107 (2007).

[14] M. Jean-Baptiste, *Journal of Physics D: Applied Physics* **46**, 143001 (2013).

[15] X. Chen, X. Zhu, W. Xiao *et al.*, *ACS Nano* **9**, 4210 (2015).

[16] M. Isasa, A. Bedoya-Pinto, S. Vélez *et al.*, *Applied Physics Letters* **105**, 142402 (2014).

[17] M. Valvidares, N. Dix, M. Isasa *et al.*, *Physical Review B* **93**, 214415 (2016).

[18] M. Isasa, S. Vélez, E. Sagasta *et al.*, *Physical Review Applied* **6**, 034007 (2016).

[19] H. Wu, Q. Zhang, C. Wan *et al.*, *IEEE Transactions in Magnetics* **51**, 4100104 (2015).

[20] T. N. Takeshi Tainosho, Jun-ichiro Inoue, Sonia Sharmin, Eiji Kita, *AIP Advances* **7**, 055936 (2017).

[21] Y.-T. Chen, S. Takahashi, H. Nakayama *et al.*, *Physical Review B* **87**, 144411 (2013).

[22] S. Y. Huang, X. Fan, D. Qu *et al.*, *Physical Review Letters* **109**, 107204 (2012).

[23] C. Tang, P. Sellappan, Y. Liu *et al.*, *Physical Review B* **94**, 140403 (2016).

[24] X. Zhou, L. Ma, Z. Shi *et al.*, *Physical Review B* **92**, 060402 (2015).

[25] G. Kresse and J. Furthmüller, *Phys. Rev. B* **54**, 169 (1996).

[26] J. P. Perdew, K. Burke, and M. Ernzerhof, *Phys. Rev. Lett.* **77**, 3865 (1996).

[27] A. Walsh, S.-H. Wei, Y. Yan *et al.*, *Phys. Rev. B* **76**, 165119 (2007).

# Supplementary Materials for:

# Magnetic Proximity Effect in Pt/CoFe$_2$O$_4$ Bilayers

Walid Amamou,[1*] Igor V. Pinchuk,[2*] Amanda Hanks,[3,4] Robert Williams,[3] Nikolas Antolin,[4] Adam Goad,[2,5] Dante J. O'Hara,[1] Adam S. Ahmed,[2] Wolfgang Windl,[4] David W. McComb,[3,4] and Roland K. Kawakami[1,2**]

[1]*Materials Science and Engineering, University of California, Riverside, CA 92521*
[2]*Department of Physics, The Ohio State University, Columbus, OH 43210*
[3]*Center for Electron Microscopy and Analysis, The Ohio State University, Columbus, OH 43210*
[4]*Department of Materials Science and Engineering, The Ohio State University, Columbus, OH 43210*
[5]*Department of Physics, University of Maryland, Baltimore County, MD 21250*

## 1. Methods

Samples are grown in a molecular beam epitaxy (MBE) chamber with base pressure of ~1x10$^{-10}$ torr. MgO(001) substrates (10 mm × 10 mm × 0.5 mm, double-sided polished from MTI) are rinsed in de-ionized water, loaded into the MBE chamber, annealed at 600 °C for 30 minutes, and smoothed by subsequent deposition of a ~5 nm electron-beam evaporated MgO buffer layer grown at 350°C at a rate of ~1 Å/min. Growth temperatures are measured by a thermocouple placed near the substrate and deposition rates are measured by a quartz crystal monitor. CFO films are deposited at ~4 Å/min in an oxygen partial pressure of 5×10$^{-7}$ torr by co-depositing elemental Co (99.99%, Alfa Aesar) and Fe (99.99%, Alfa Aesar) from thermal effusion cells. The substrate temperature is maintained at 200 °C during CFO growth and *in situ* reflection high energy electron diffraction (RHEED) is used to monitor the sample surface throughout the growth and annealing process. CFO films are then cooled to room temperature and capped with either Pt, Pt/Cu or Cu. Pt films are deposited at ~0.06 Å/min using an electron beam source while Cu films are grown at 1 Å/min using a thermal effusion cell. The described heterostructures are deposited without breaking UHV conditions in order to preserve the quality of the Pt/CFO and Cu/CFO interfaces. Magnetization measurements are performed using a Quantum Design 14 Tesla Physical Properties Measurement System (PPMS) with a vibrating sample magnetometer (VSM) module. The samples are patterned into Hall bars (width W=100 μm, length L = 800 μm) for subsequent DC magnetoresistance and Hall measurements. DC transport measurements are obtained in the same PPMS using resistivity mode. For all Hall measurements, we apply a DC current ($I$ = 20 μA) and measure the transverse voltage $V_{xy}$ as an out-of-plane magnetic field is swept. The Hall resistivity is given by $\rho_{xy} = (V_{xy}/I)t$, where $t$ is the thickness of the Pt channel. Angle-dependent magnetoresistance measurements are performed by placing Hall bars into a constant magnetic field of 10 Tesla and rotating the sample stage.

**2. Room temperature measurements**

In order to test for the presence of MPE at room temperature, we repeat the Hall measurement and angle-dependent magnetoresistance measurements at 300 K with the results shown in Figures S1. In Figure S1a, we show the Hall resistivity after subtraction of the OHE linear background. Interestingly, the Hall signal still shows a nonlinear, hysteretic signal with a smaller coercivity and remanence than observed at 5 K. Figure S1b shows the angle-dependent β-scan of longitudinal resistance, which is sensitive to SMR. The SMR ratio $\Delta\rho_{xx}/\rho_0$ has nearly the same magnitude at 300 K as at 5 K (Fig. 3b of the main text), demonstrating negligible temperature dependence similar to previous reports in Pt/CFO bilayers [1]. Figure S1c shows the angle-dependent γ-scan of the longitudinal resistance, which is sensitive to AMR and OMR. The most notable feature is the opposite polarity of the angle-dependence in the γ-scan at 300 K compared to 5 K (Fig. 3c of the main text). Because the 5 K scan has the same polarity as OMR, there are two possible scenarios: (1) only OMR is present and there is no induced ferromagnetism in Pt (i.e. no AMR signal), or (2) AMR is present due to induced ferromagnetism in Pt, but the OMR signal dominates over the AMR signal. Therefore, the negative polarity of the γ-scan at 300 K means that it cannot provide conclusive evidence for MPE at room temperature. This contrasts with the case at 5 K where the positive polarity of the γ-scan which requires the presence of AMR and induced ferromagnetism to explain the observed signal.

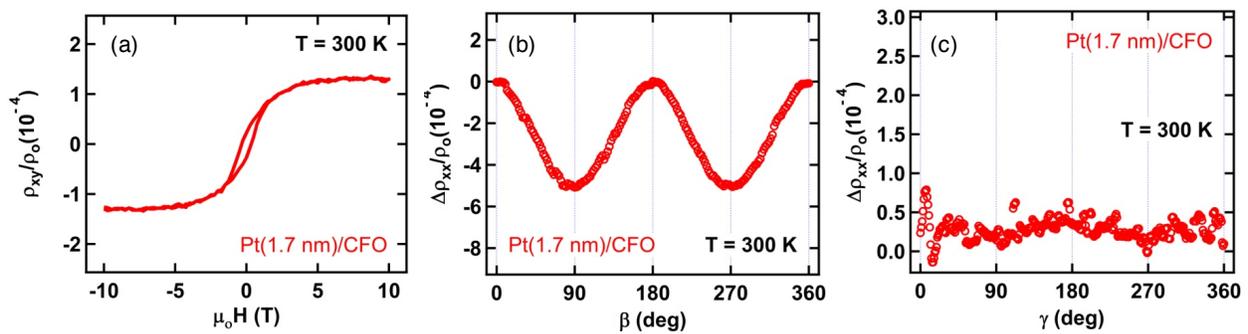

**FIG. S1.** Room temperature measurements. (a) Anomalous Hall effect for Pt/CFO. The linear OHE background has been subtracted. (b) Angle-dependent β-scan of Pt/CFO, showing SMR. (c) Angle-dependent γ-scan of Pt/CFO, exhibiting negative polarity of the MR signal.

## 3. Orbital density of states DFT calculation

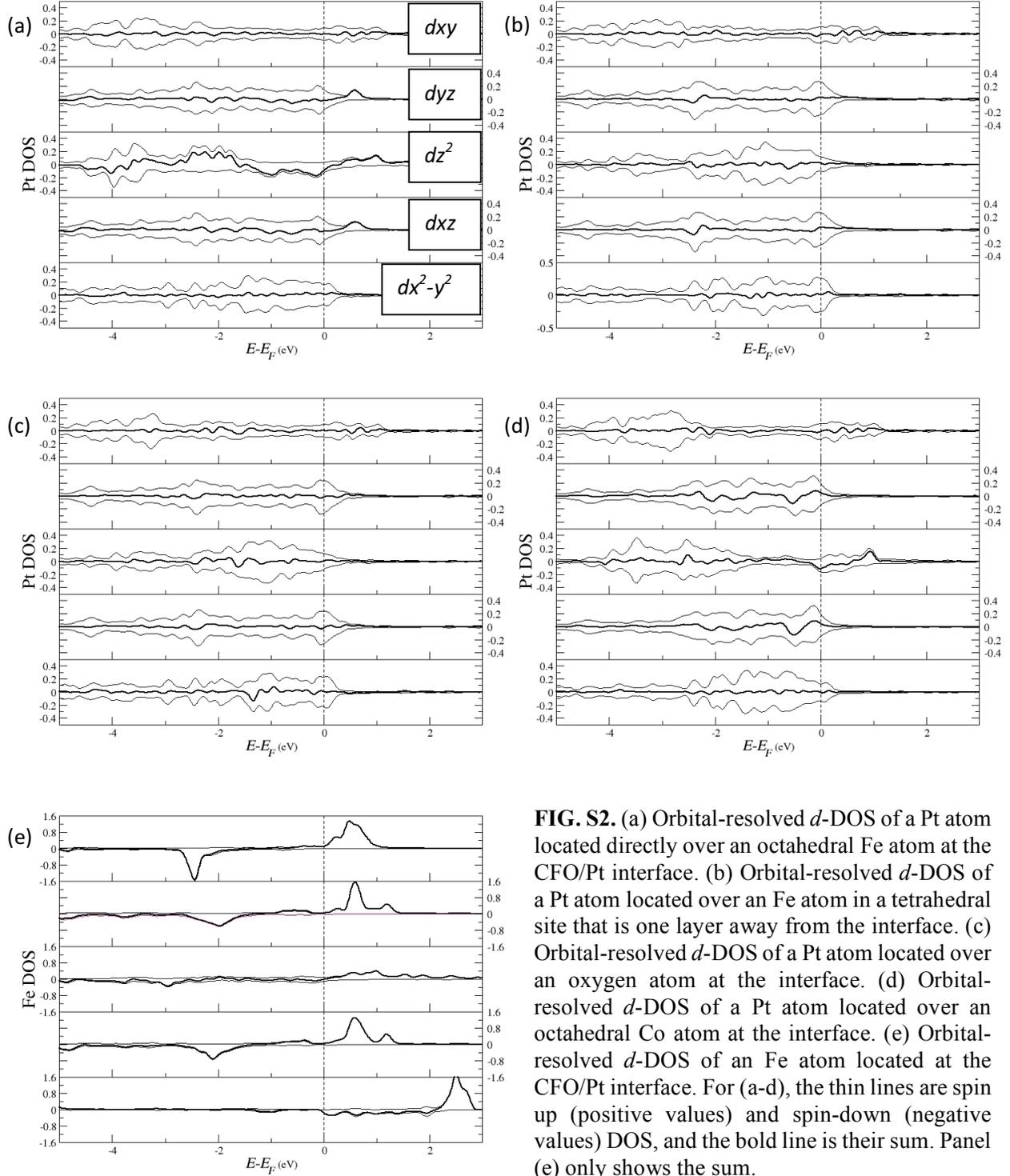

**FIG. S2.** (a) Orbital-resolved *d*-DOS of a Pt atom located directly over an octahedral Fe atom at the CFO/Pt interface. (b) Orbital-resolved *d*-DOS of a Pt atom located over an Fe atom in a tetrahedral site that is one layer away from the interface. (c) Orbital-resolved *d*-DOS of a Pt atom located over an oxygen atom at the interface. (d) Orbital-resolved *d*-DOS of a Pt atom located over an octahedral Co atom at the interface. (e) Orbital-resolved *d*-DOS of an Fe atom located at the CFO/Pt interface. For (a-d), the thin lines are spin up (positive values) and spin-down (negative values) DOS, and the bold line is their sum. Panel (e) only shows the sum.

In Figure 4a of the main text, the large error bar for the average magnetic moment per Pt atom in Pt layer #1 (interfacial layer) reflects the site-specific variation of the induced magnetic moment. To investigate the nature of the induced moments, we examine the orbital density of states of various Pt atoms in layer #1. Figure S2a shows the *d*-DOS of a Pt atom located directly over an octahedral Fe atom. Here, we observe substantial spin asymmetry of the density of states (bold line) for the $d_{z^2}$ orbitals, but much weaker spin asymmetry for the other orbitals. This indicates a strong magnetic proximity effect for Pt atoms on top of octahedral Fe atoms, due to induced spin polarization in the $d_{z^2}$ orbitals. Figure S2b shows the *d*-DOS of a Pt atom located over a Fe atom in an tetrahedral site, which lies one layer away from the interface. As is clear from the calculation, none of the Pt *d*-orbitals show any substantial spin asymmetry. Similarly, the Pt atoms on top of an O atom (Figure S2c) and on top of a Co atom (Figure S2d) show very little spin asymmetry of the DOS for all Pt *d* orbitals. Thus, the induced magnetism in Pt is due to spin polarization of the $d_{z^2}$ orbitals for Pt atoms on top of an octahedral Fe atom. To investigate which *d*-orbitals of the Fe atom contribute to the magnetic exchange coupling, we consider the *d*-DOS of the octahedral Fe atoms located at the CFO/Pt interface, as shown in Figure S2d. Notably, only the Fe $d_{xy}$, $d_{yz}$, and $d_{xz}$ orbitals show the strong spin asymmetry and magnetic moments, which suggests that magnetism is primarily induced in the $d_{z^2}$ orbital of the Pt atoms by moments in the $d_{xy}$, $d_{yz}$, and $d_{xz}$ orbitals of Fe atoms no more than one layer removed from the CFO/Pt interface.